\documentstyle[12pt,aaspp]{article}
\newcommand\etal{{\it et~al.}}

\begin{document}

\title{VLBI Observations of the Gravitational Lens System 0957+561}

\centerline{\it To appear in the Astronomical Journal, 1995.12\,.}
\centerline{\it Submitted 1995.07.06, accepted 1995.08.17\,.}

\author{R.~M. Campbell\altaffilmark{1}, J. Leh\'ar,}
\affil{Center for Astrophysics, 60 Garden Street, Cambridge, MA 02138 \\
Electronic mail:  \{rcampbell, jlehar\}@cfa.harvard.edu}

\author{B.~E. Corey,}
\affil{Haystack Observatory, Off Route 40, Westford, MA 01866 \\
Electronic mail:  bec@wells.haystack.edu}

\author{I.~I. Shapiro, and E.~E. Falco}
\affil{Center for Astrophysics, 60 Garden Street, Cambridge, MA 02138 \\
Electronic mail:  \{ishapiro, falco\}@cfa.harvard.edu}

\altaffiltext{1}{present address: Ionospheric Effects Division,
Geophysics Directorate, Phillips Laboratories, Hanscom~AFB, MA 01731}


\begin{abstract}

We present hybrid maps of the A and B~images of 0957+561
from each of
four sessions of 6\,cm VLBI observations that span
the six-year interval 1987--1993.
The inner- and outer-jets are clearly detected
and confirm the structures reported previously.
There is no evidence of change in the separation between the core and
inner-jet components,
so the prospect of measuring the time delay
using differential proper motions is not promising.
The flux density in the core of each image peaked between 1989 and 1992.
{}From the variation in these flux densities,
we obtain a time-delay estimate of $\sim1{\rm\,yr}$.

\end{abstract}

\newpage

\section{Introduction}

The first gravitational lens system to be discovered, 0957+561
(Walsh, Carswell, \& Weymann 1979),
remains the most extensively studied and discussed.
A major cause of this attention is
the prospect of obtaining an estimate of $H_0$ directly from a
cosmologically distant source, bypassing the many calibration-sensitive
rungs of the ``cosmic distance ladder.''  To use a gravitational lens system
for this purpose, we require a determination of
the mass distribution within the lens
({\it e.g.,} Falco, Gorenstein, \& Shapiro 1991)
and the ``time delay,'' $\Delta\tau$, the difference
in propagation times from source to observer
via two images (Refsdal 1966).

Efforts to determine $\Delta\tau$ have concentrated on examining correlations
between the light curves of the A and B~images (in the sense that $\Delta\tau
> 0$ implies that image A precedes image B).
However, different investigators have obtained different results.
Optical light-curves have yielded $\Delta\tau\approx1.0{\rm\,yr}$
(Schild~\& Thomson 1995 and references therein;
Vanderriest \etal\ 1989),
while the VLA light-curves have yielded
$\Delta\tau\approx1.5{\rm\,yr}$ (Leh\'ar \etal\ 1992).
This discrepancy led to the reanalysis of various
portions of these data (Press, Rybicki, \& Hewitt 1992;
Pelt \etal\ 1994;
Thomson~\& Schild 1995).  These studies illustrate
the difficulties in interpreting the light-curve data.

   Campbell \etal\ (1994, hereafter C94) discussed
two sessions of
6\,cm VLBI observations.
Their primary intent was to detect proper motion between
the core and inner-jet components of each image.
Such proper motion would provide an estimate of $\Delta\tau$,
independent of the light-curve analysis.
Unfortunately, no significant proper motion was found.  Turning instead
towards further constraint of
the mass distribution within the lens through investigation of gradients
in the relative magnification matrix for the two images,
we carried out two additional sessions of VLBI observations,
separated by the presumed $\Delta\tau$ of 1.5\,yr
(Press \etal\ 1992).
Save for any effects of microlensing,
comparison of the two images at the same ``source epoch'' would
isolate the lensing distortions from any intrinsic quasar variability,
and thereby provide better constraints on the mass distribution.

\section{Observations}

We have made simultaneous 6\,cm observations of the A and B~images
of 0957+561 in four sessions (see Table~1).
C94 describe results from the first two sessions more fully.
The 0957+561 images were observed in 12~minute scans
at intervals of about 30~minutes.
A few compact sources were also observed during each session
to aid fringe finding.
The reference frequency was 4.983\,GHz and the recorded
bandwidth $B$ was 56\,MHz, except at the VLA where $B=48$\,MHz as limited
by the front-end filter.
We relied on radiometry information supplied by each station
to provide the initial flux-density calibration.

For the first two observing sessions (1987 and 1989),
each station  used the Mk\,III recording format
(Rogers \etal\ 1983) in mode~A
(14 independent video channels).
In the third session (1992), we incorporated
some of the new VLBA antennas near the VLA to provide more short baselines.
However, the Los~Alamos station produced no useful data,
due to an equipment failure.
Whenever they participated, VLBA stations
used the Mk\,III recording format in mode~B (7 video channels, $B=28$\,MHz).
This resulted in a $\sim30\%$ loss in sensitivity
for baselines involving at least one of those stations.
There were further sensitivity losses in the fourth session (1993):
the Hancock VLBA station was substituted
for the more sensitive Haystack antenna,
and, more significantly, the VLA replaced their Mk\,III acquisition system
with a VLBA acquisition rack, resulting in a similar bandwidth reduction.

     Data tapes were correlated on the Mk\,III\,A processor
at Haystack Observatory.
Specific correlation procedures are discussed in C94.
Baselines involving only a single antenna
at each station were correlated twice,
once for the coordinates of each image.
The VLA and Westerbork could not observe both images simultaneously,
so these stations observed each image on alternate scans.
Baselines formed exclusively from the set of stations comprising L, K, O,
and VLBA antennas (see Table 1 for identifications) were not
sensitive enough to detect the images.
Detections of the B~image on baselines such as G--VLBA were marginal.
Following data export from the Haystack correlator,
we used the Caltech VLBI Software Package (Pearson 1991)
to perform all editing and calibration.

\section{Hybrid Maps}

We produced hybrid maps of 0957+561 using the DIFMAP software
in the Caltech VLBI package
(Shepard, Pearson, \& Taylor 1994).
We initially calibrated the station phases to a point source model,
with a flux density of 10\,mJy for the A~image and 7\,mJy
for the B~image.
Eight cycles of CLEAN mapping and self-calibration then followed.
In each cycle, we started with a new set of CLEAN components,
and broke the CLEANing into four runs,
increasing the number of iterations in each successive run.
For the first two cycles, we used a small CLEAN window
which enclosed only the core and the inner-jet components.
In the next two cycles, we included a second window
which covered the brightest part of the outer-jet
(Jet~2~in Gorenstein \etal\ 1988).
In the third set of two cycles, we used six windows
to cover the entire VLBI structure of each image.
In the final two cycles, we used the GSCALE routine,
which adjusts each station gain by a constant factor
to improve the fit of the CLEAN model to the visibilities.

Using this procedure, we produced maps for all four sessions;
Table~2 shows some of the results.
The self-calibration procedure evaluates closure phases
for each individual scan,
so those visibilities whose baselines did not form
part of a closure triangle at a given time were deleted.
The residual map $rms$ values were $\sim0.04{\rm\,mJy/beam}$
for all sets of observations except those from 1993 (our least sensitive
array, see \S 2 above),
where the $rms$ values were $\sim0.06{\rm\,mJy/beam}$.
Table~2 lists for each observing session the reduced chi-square,
which DIFMAP computes by comparing the observed visibilities
to those predicted from the CLEAN model.
We have integrated the flux density in a window which
encloses only the core and inner-jet components,
and we list these as well in Table 2,
with standard errors derived from the map $rms$.
We also list the $rms$ of the GSCALE corrections,
as a conservative estimate of the absolute flux-density calibration.
The A and B~image GSCALE corrections were consistent
to $\sim4\%$ on average, and their ratio
did not differ from zero by more than $\pm7\%$ in any case.

We show the hybrid maps of the inner regions of each image
in Figure~1.
All the maps show the same basic structure seen in C94.
In particular, there is no clear evidence
of any change with time in the separation between the core
and inner-jet components for either image.
We are thus not likely to measure the time delay from
differential proper motions in the near future.

The outer-jet is detected in the data from each of the four sessions,
and shows similar structure in each.
This structure is less reliably determined for the 1989 and 1993 sessions,
however, due to poorer data quality (see \S 2 above; C94).
Figure~2 shows the 1992 hybrid maps,
convolved with a circular beam of ${\rm FWHM=6\,mas}$
to emphasize the more extended outer-jet structure.
The outer-jet compares very well with
that in the $18{\rm\,cm}$ maps of Garrett \etal\ (1994),
and is consistent with the general morphology of the elliptical Gaussian
models used in C94.
We will report elsewhere
on determining the image magnification gradient
from the outer-jet structures.

\section{Estimation of the Time Delay}

The flux densities in the cores of the A and B~images vary,
and both show a clear peak between 1990 and 1992 (see~Table~2).
We can obtain from these flux-density measurements a
crude estimate of $\Delta\tau$
and the ratio $R$ of the magnification of the core of the B~image
to that of the A~image.
Although the time sampling is very sparse, we can make this estimate
by assuming that the core brightness varies smoothly.

We determined $\Delta\tau$ and $R$ simultaneously by fitting a polynomial
to the combined A~image and (shifted) B~image light curves
({\it e.g.,} Leh\'ar \etal\ 1992).
Using our core flux-densities (Table~2),
we considered a two-dimensional grid of shifts,
covering $-1.5\,{\rm yr} < \Delta\tau < 3.5\,{\rm yr}$ and $0.4<R<0.9$\,.
For each grid point, we obtained a least-squares polynomial fit to
the combined light-curves using SVDFIT (Press \etal\ 1989),
and the total $\chi^2$.
We set the standard errors of $\Delta\tau$ and $R$ as
the range over which
$\chi^2-\chi^2_{\rm min}<1$, where
$\chi^2_{\rm min}$ denotes the minimum value obtained for $\chi^2$.

Using the map $rms$ to set the flux-density errors, we found that
a fourth-order polynomial was the lowest order required to model
the observed shape of the light-curve,
since a third-order polynomial led to
a 100-fold increase in $\chi^2_{\rm min}$.
Thus, with five polynomial parameters,
two shift parameters ($\Delta\tau$,$R$),
and eight flux-density constraints,
only one degree of freedom remains for each fit.
We obtained $\Delta\tau=0.88\pm0.13{\rm\,yr}$ and $R=0.664\pm0.007$,
with $\chi^2_{\rm min}=0.31$.
The upper panel of Figure~3 shows the combined light-curve data with
the best-fit polynomial.

A very conservative error estimate is
the $rms$ of the GSCALE station corrections (see~Table~2),
because the flux-density calibration is not likely to be wrong
by more than the station radiometry errors.
Note that radiometry errors for each station
are likely to be correlated between sessions.
A third-order polynomial suffices to fit
the core flux-densities with these errors,
giving two degrees of freedom.
The resulting time delay and flux-density ratio are
$\Delta\tau=0.5\pm 0.5{\rm\,yr}$ and $R=0.67\pm0.08$,
with $\chi^2_{\rm min}=1.86$.
The lower panel of Figure~3 shows the corresponding best-fit polynomial.
$\chi^2_{\rm min}$ doubled if we used a second-order polynomial to fit
the flux-density variations.

Our image magnification parameter $R$
differs considerably from the core magnification
found by other investigators
(Conner, Leh\'ar, \& Burke 1992, and references therein).
However, there are reasons why our result could be in error.
Most importantly, the A and B~images have different
effective baseline sampling, because they are magnified by different factors.
The inner-jet of the B~image is longer than that of the A~image,
so some of its flux density may be resolved out by the VLBI sampling.
Note also that the B image is dimmer than the A image;
the consequently greater number of non-detections of the B image
results in
a sparser ({\it u--v\/}) sampling for it than for the A image,
independent of any relative magnification effects.

   In any event, the estimated time delay is less sensitive to
considerations concerning the ({\it u--v\/}) sampling,
since $\Delta\tau$ is determined only from the varying portion of the
flux density in the core.
Given that no structural variations are seen in the inner jet,
its brightness is not likely to vary on a timescale of a few years.
Thus, any inner-jet flux density which we resolve out in the B~image
should only affect $R$ and not $\Delta\tau$.
Since the flux density of the B~image decreased sharply in 1993,
our light-curve favors the shorter ``optical'' delay
of $\sim1{\rm\,yr}$
over the longer ``radio'' delay of $\sim 1.5$\,yr,
but our formal errors are too large to exclude either case.
The variation of the flux density in the inner regions of the images
from the first two sessions are consistent
with the variations observed by the VLA monitoring program
(C94, Leh\'ar \etal\ 1992).
The subsequent decrease in flux density observed for both images
during the last two VLBI sessions should also have been observed
with the VLA.
Because these latter data contain much more closely spaced samples,
and since their calibration should be more reliable,
the VLA monitoring program should yield
a more precise estimate of $\Delta\tau$ from this feature.

\acknowledgements

We thank participating observatories and their staffs for supporting
our experiments, especially the Haystack correlator staff for ensuring
expeditious processing of the 1993 data.  Operation of the
Mk\,III correlator at Haystack Observatory was funded by the NSF through
the Northeast Radio Observatory Corporation.  We acknowledge support from
NASA grant NGT-50663 (RMC) and NSF grants AST89-02087 and AST93-03527
(RMC, JL, and IIS).



%
\newpage
\begin{picture}(367,475)
\put(0,0){\includegraphics{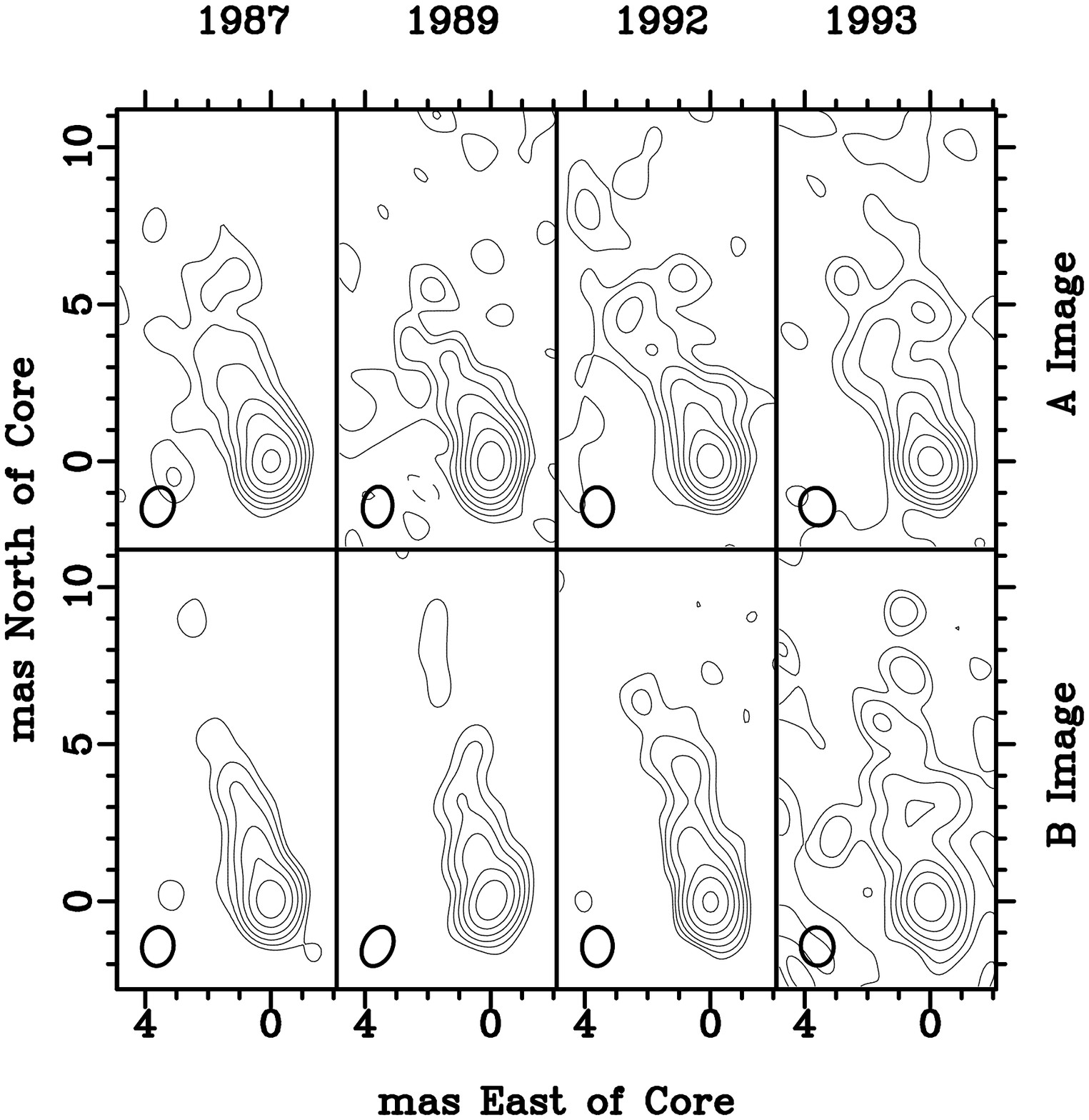}}
\end{picture}
\\
\noindent
{\bf Figure 1:}
Hybrid maps of the core
and inner-jet of 0957+561 A and B,
for all four sessions of 6\,cm observations.
The $\sim1{\rm\,mas}$ diameter restoring beam is shown as an ellipse in
the lower left of each field,
and was chosen to represent the angular resolution of each observation,
using the ``uniform'' baseline weighting scheme (Shepard \etal\ 1994).
The pixel size is $0.1$\,mas, and the map contours increase by
factors of two from $0.125{\rm\,mJy/beam}$ to $8{\rm\,mJy/beam}$.

%
\newpage
\begin{picture}(367,475)
\put(0,0){\includegraphics{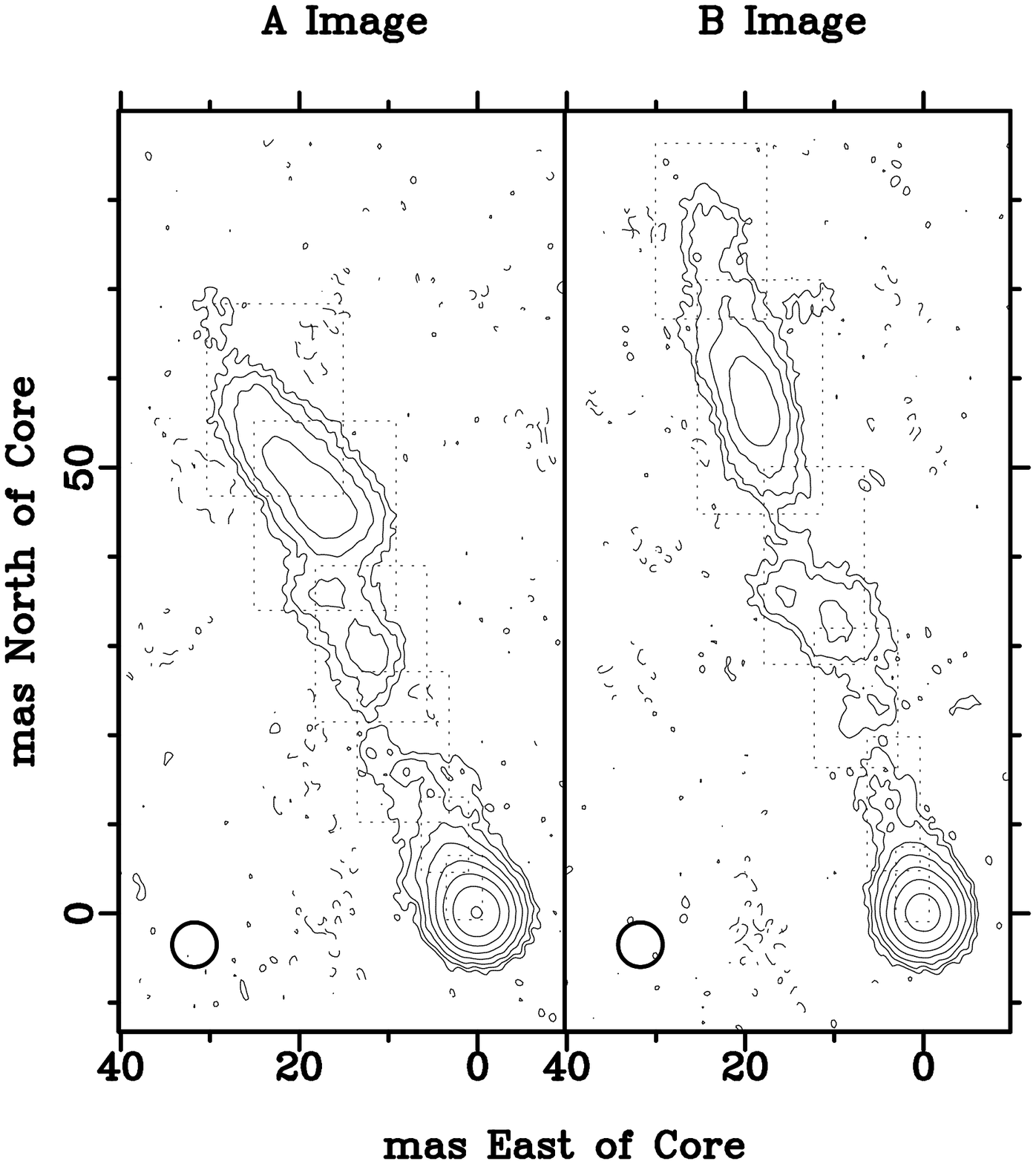}}
\end{picture}
\\
\noindent
{\bf Figure 2:}
Hybrid maps
of 0957+561 A and B, from the 1992 6\,cm
observations.  We convolved
the clean components with a circular, ${\rm FWHM} = 6{\rm\,mas}$
beam to emphasize the outer-jet structure.
The six CLEAN windows are outlined with dotted lines,
and the restoring beam is shown in the lower left of each field.
The pixel size is $0.25$\,mas, and the map contours increase by
factors of two from $0.125{\rm\,mJy/beam}$ to $16{\rm\,mJy/beam}$.

%
\newpage
\begin{picture}(367,475)
\put(0,0){\includegraphics{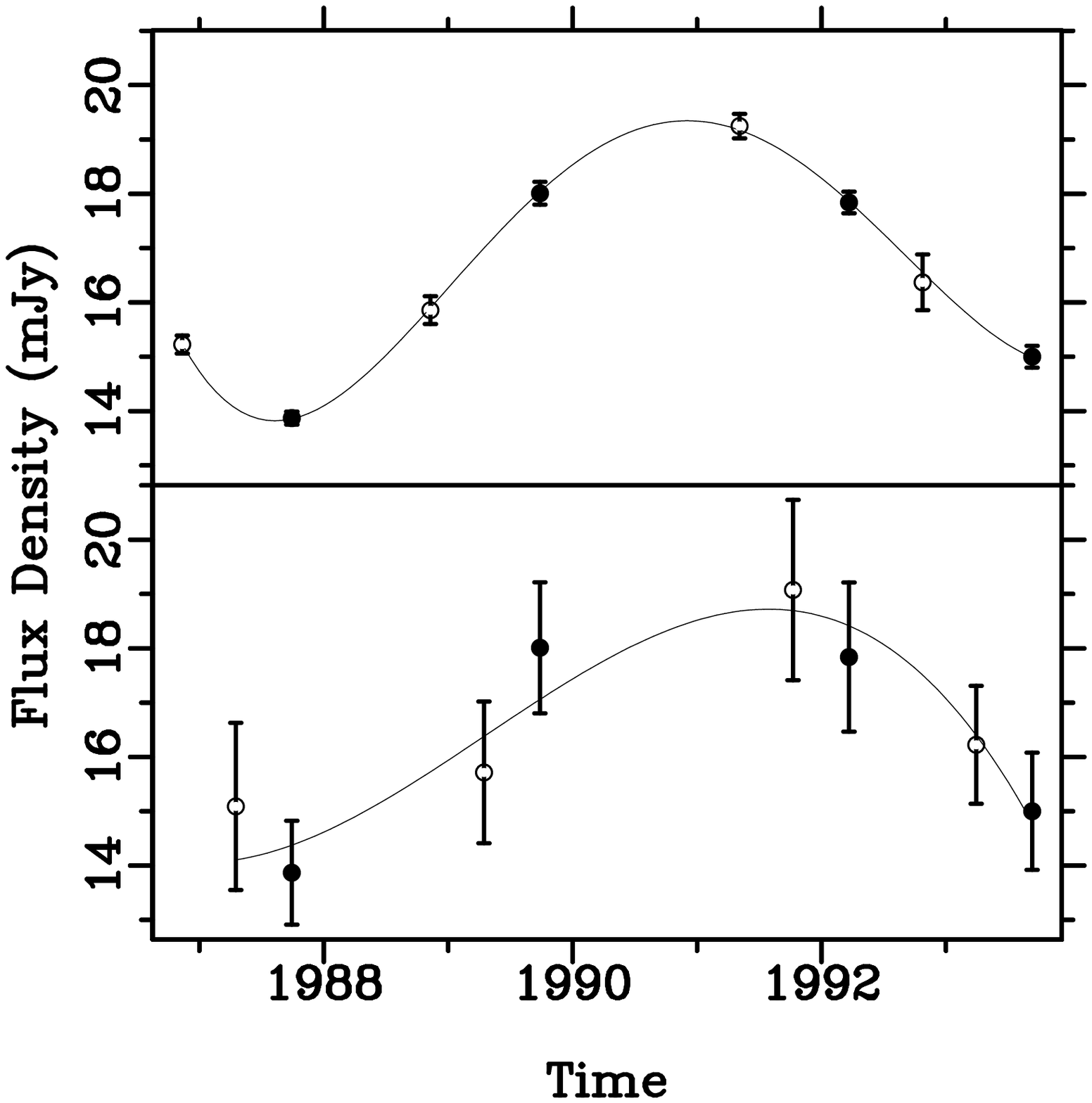}}
\end{picture}
\\
\noindent
{\bf Figure 3:}
Combined VLBI
light curves for the core of 0957+561.
The observed A-image flux densities are shown as filled circles.
The B-image data have been shifted by the least-square
estimates of $\Delta\tau$ and $R$,
and are shown as empty circles.  We also show the
least-squares polynomial light curves.
The upper panel shows the result of analysis
using the map $rms$ flux-density errors ({\it i.e.,} we fit
a fourth-order polynomial light curve),
and the lower panel shows the result of analysis using the GSCALE
error estimate ({\it i.e.,} we fit a third-order polynomial
light curve).

\newpage
%
\begin{table*}[h]
\caption{Summary of Observations}
\begin{tabular}{ccccl}
\tableline
\tableline
& & & \#Baselines & \nl
Start Date &
Start UT &
Duration &
(Image A,B) &
Stations\,\tablenotemark{a,b} \nl
\tableline
1987.09.28 & 09:30 & 13 hr & 18,12 & L--B--W--K--G--Y--O \nl
1989.09.26 & 09:00 & 15 hr & 14,13 & B--W--K--G--Y--O \nl
1992.03.21 & 21:00 & 13 hr & 15,15 & L--B--K--G--Y--Kp--Ov \nl
1993.09.10 & 09:30 & 15 hr & 17,17 & L--B--Hn--G--La--Y--Kp--Ov \nl
\end{tabular}
\tablenotetext{a}{
 VLBI Network: B=Bonn, G=Green Bank 43m, K=Haystack, }
\tablenotetext{~}{
   L=Medicina, O=Owens Valley 40m, W=Westerbork, Y=VLA}
\tablenotetext{b}{
 VLBA Antennas: Hn=Hancock, Kp=Kitt Peak, La=Los Alamos, Ov=Owens Valley}
\end{table*}

\vfill

\begin{table*}[h]
\caption{Hybrid Mapping Results}
\begin{tabular}{rrrrrr}
\tableline
\tableline
 Observation &
 \#Visibilities  &
 Core Region &
 Scale &
 Reduced
\nl
 Date:Image &
 Total:Deleted\tablenotemark{a} &
 Flux Dens.\tablenotemark{b} &
 Error\tablenotemark{c} &
 $\chi^2$
\nl
\tableline
1987:A &  692:48 & $13.87\pm0.12$ &  6.9\% & 0.86 \nl
1989:A &  534:96 & $18.01\pm0.21$ &  6.7\% & 0.98 \nl
1992:A &  855:80 & $17.84\pm0.20$ &  7.7\%\tablenotemark{d} & 0.93 \nl
1993:A &1068:196 & $15.00\pm0.20$ &  7.2\% & 1.13 \nl
\nl
1987:B & 529:145 & $10.11\pm0.11$ & 10.2\% & 0.68 \nl
1989:B & 411:183 & $10.53\pm0.17$ &  8.3\% & 0.88 \nl
1992:B & 785:104 & $12.78\pm0.15$ &  8.7\%\tablenotemark{d} & 0.93 \nl
1993:B &1028:194 & $10.87\pm0.34$ &  6.7\% & 1.48 \nl
\end{tabular}
\tablenotetext{a}{\,visibilities were deleted if either antenna was
                  not part of any closed loop of baselines}

\tablenotetext{b}{\,in mJy; error from the $rms$ of the CLEAN map}

\tablenotetext{c}{\,the $rms$ of station gain corrections in GSCALE}

\tablenotetext{d}{\,excludes the VLBA antennas, whose $T_{sys}$ values
                    were $>20\%$ higher than expected}

\end{table*}

\end{document}